\documentclass[12pt,draft]{iopart}

\newcommand{\pa}{\partial}

\usepackage{amssymb}
\usepackage{iopams}
\usepackage{footnote}

\begin{document}

\title[A theorem on 1/f %
noise]{Quantum Brownian motion \\
and a theorem on fundamental 1/f noise}

\author{Yu E Kuzovlev}

\address{Donetsk Institute
for Physics and Technology of NASU, %
ul. R.Luxemburg 72, 83114 Donetsk, Ukraine}
\ead{kuzovlev@fti.dn.ua}

\begin{abstract}
We consider quantum Hamiltonian systems composed of mutually interacting %
``dynamical subsystem'' with one or several degrees of freedom %
and `thermostat'' with arbitrary many degrees of freedom, %
under assumptions that the interaction %
ensures irreversible behavior of the dynamical subsystem, %
that is finite diffusivities  %
of its coordinates in thermodynamically equilibrium state and finite %
drift velocities and mobilities in non-equilibrium steady state in presence of %
external driving forces. It is shown that, nevertheless,  %
regardless of characteristics of the interaction, the diffusivity %
and mobility have no certain values but instead vary from %
one observation to another and undergo 1/f-type or %
flicker-type low-frequency fluctuations.
\end{abstract}


\vspace{2pc} \noindent{\it Keywords}\,: %
Dynamical foundations of kinetics, %
Quantum Brownian motion, Quantum transport, %
Quantum kinetic equations, %
Mobility fluctuations, 1/f noise, %
Low-frequency noise, Flicker noise


\section{Introduction}

Nearly 30 years ago in \cite{pr1,pjtf} and then %
in \cite{bk1,bk2,bk3,pr2} an original explanation %
of electronic 1/f-noise was suggested by Prof. %
G.\,N.\,Bochkov and me and putted in phenomenological %
statistical model which easy produces correct estimates %
of the noise level and directly connects to rigorous %
statistical mechanics (for later looks at this model %
see \cite{i2,p1008}). Key idea of our explanation was %
that own (thermal) fluctuations in rates of various %
transport processes in many-particle systems %
do not cause a ``back reaction'' and therefore have %
no definite ``relaxation times''. Hence, that are scaleless %
fluctuations, with power-law low-frequency spectrum %
(which thus has no relation to folklore ``composition of lorentzians''). %

In \cite{pr1,pjtf,bk1,bk2,bk3,pr2} this %
principal idea was exploited in specific terms of electric charge transport %
and ``Brownian motion'' of charge carriers %
(one may see also \cite{kmg,july,oct,mang}). %
But, clearly, it can be applied also to to different processes %
in large variety of physical systems (in most wide sense of the word %
``physical'' \cite{i2}). %
Particularly, to Brownian motion (``random walks'') of %
a fluid particles \cite{i1,p1,jstat,tmf,ig,lpro,yuk}, as well as to %
collective fluctuations and non-equilibrium processes in fluids %
\cite{jstat,tmf} and solids \cite{i3}. %

Simultaneously, the works \cite{i1,i3} and later %
\cite{p1,jstat,tmf,ig,lpro} and \cite{oct}, %
in fact,  presented substantiations of the mentioned %
key idea on the base of rigorous statistical mechanics, %
and confirmed the sentence from \cite{bk3} that %
``1/f-noise is a kind of tribute to be paid to the dynamics for %
dissipation and irreversibility properties of physical systems'' %
(my translation from Russian original). %

This sentence fully agrees with results of the deep %
N.\,Krylow's investigation \cite{kr} (discovered by me %
when preparing \cite{i1}) where it was shown that  %
in many-particle statistical mechanics, firstly,    %
relative frequencies of observable events can be different %
at different system's phase trajectories (experiments), %
regardless of duration of observations and time averaging %
(see characteristic mental example in introductions in \cite{tmf} %
and \cite{p1} and more examples in \cite{i2}). %
Therefore, generally one has no rights to presume any \,{\it a priori}\, %
certain (even let unknown numerically) ``probabilities'' of events. %
Secondly (as the consequence), it is possible that %
statistical correlations do exist even betwen events %
what are independent in physical (cause-and-consequence) %
sense.

In \cite{i1,i3,p1,jstat,tmf} I demonstrated that %
real ``fundamental'' 1/f-noise (i.e. such 1/f-noise %
whose spectrum has no saturation at zero frequency) is just  %
manifestation of the Krylov's uncertainty of events' probabilities %
(additional explanations can be found in \cite{i2} %
and introductory or discussion-resume sections in %
\cite{p1008,kmg,ig,lpro,dvr} and \cite{oct}). %

Unfortunately, the enumerated works, - including the %
Krylov's book, - yet has not excited any scientific-community's %
response except neglect or aversion %
\footnote{\,

Thus showing that progress in fundamental science %
also undergoes fundamental 1/f-type fluctuations. %
For example, manuscripts \cite{jstat} were rejected by JSTAT %
and JSP, respectively, \cite{tmf} by Physica A, %
\cite{ig} by JSP, EPL and JMP, \cite{p1008} and \cite{july}-\cite{oct} by JETP, %
\cite{dvr} by PRE, JPA and JSP, etc., %
in all cases without meaningful or even any motivation. %
} %
. Interestingly, such aversion was pointed out and %
commented already by Krylov himself \cite{kr}, %
as product of common prejudices  educated %
by standard ``probability-theoretical'' view at physical world %
and its statistical description %
\footnote{\,

The modern notion of ``dynamical chaos'' %
still has not changed this situation, since, instead of investigating %
specificity of the chaos in (infinitely-) many-particle Hamiltonian systems, - %
where it naturally creates 1/f-noise \cite{tmf,i3} %
and thus ``kinetic non-ergodicity'' \cite{i2,tmf}, - scientists try %
to  squeeze it in ``Procrustean bed'' formed with %
``Marcovian partitions'', ``Bernoullian flows'' %
and primitive stochastic processes with \,{\it a priori}\, %
introduced ``probabilities of elementary events''. Although %
there is also concept of such ``K-flows'' what %
can not be re-coded to Bernoullian flows \cite{orn}. %
} %
. %
Hence, we have to continue activity in this field as far as possible. %

\,\,\,

In this paper our aim is to illustrate the aforesaid %
ideology (and once again confirm its correctness) %
by example of simple Hamiltonian microscopic model of quantum Brownian motion. %
More concretely, to show that if an interaction of (quantum) %
 particle with (quantum) thermostat make this particle ``Brownian'', %
that is ensures diffusive character of its chaotic motion, -  %
when mean square of its displacement grows proportionally to time, - %
then such interaction ensures also non-zero 1/f-type %
fluctuations of its diffusivity (and hence mobility). %
In other words, the latter has %
no certain value but varies from one measurement to another. %



\section{A class of systems under consideration}

Let us consider a quantum system with Hamiltonian
\begin{eqnarray}
H\,=\,H_d(q,p)+H_{b}(q)\,\,\,, \label{h}  \\ %
H_d(q,p) =\frac {p^2}{2m}\,-f\cdot q\,\,\,, \label{hd} %
\end{eqnarray}
where\, $\,H_d(q,p)\,$\, represents ``dynamical subsystem'' %
with canonical variables (coordinates and momenta) %
$q$ and $p$, whose operators satisfy standard commutation relations %
\begin{eqnarray}
[q_{\alpha}\,,p_{\beta}]= i\hbar\, %
\delta_{\alpha\beta}\,\,\,, \label{qp}
\end{eqnarray}
and\, $H_b(q)$ represents ``thermal bath'' (``thermostat'')\, %
along with interaction between it and the dynamical subsystem (DS).

Our first principal assumption is that our system is ``translationally %
invariant'' in respect to the coordinates $q$, in the sense that   %
properties of the operator $H_b(q)$, - as considered in the %
thermostat's Hilbert (phase) space, - %
do not depend on parameters $q$. %
Formally, this means that there exist such %
Hermitian operators $\Pi$, - defined in the thermostat's space, - that %
\begin{eqnarray}
H_{b}(q+a)\,=\, \exp{(-ia\Pi/\hbar )}\, %
H_{b}(q)\, \exp{(ia\Pi/\hbar )}\, \,\, \label{p} %
\end{eqnarray}
In other words, from thermodynamical point of view %
the thermostat is indifferent to $q$'s values. %

Under such assumption, one can treat variables (observables) %
$q$ and $v=p/m$ like coordinate and velocity %
of unrestricted (generally multi-dimensional) ``Brownian motion''. %

We want to consider possible statistical characteristics %
of these motion, at that basing, of course, on the von Neumann %
(quantum Liouville) evolution equation for full system's %
density matrix $\rho$, that is
\begin{eqnarray}
\dot{\rho}\, =\,[H,\rho]/i\hbar\,\equiv\, %
\mathcal{L}\,\rho\,\,\,, \label{e} %
\end{eqnarray}
with\, $[A,B]\equiv AB-BA$ and %
$\mathcal{L}$ being the Liouville super-operator.

\section{Quasi-classic representation}

It is convenient to consider Eq.\ref{e}  %
in the form most closely unifying quantum and classical treatments %
of DS's variables, that is in quasi-classical or Wigner representation. %
Simultaneously, by reasons what will be clear later, %
we want to go from the density matrix, $\rho$, to an equivalent %
characteristic function of variables of our system. %
We will make this in three steps. %

{\bf 1}.\,  %
Let  $|q\rangle$ be eigen-vectors of operators $q$, %
and let us introduce functions-operators
\begin{eqnarray}
\rho(t,x,y)\,=\, %
\langle x+y/2|\,\rho\,|x-y/2\rangle \,\,\,, \label{xy}\\
\rho(t,x,p)\,=\, \int \exp{(-ipy/\hbar)}\, %
\rho(t,x,y)\, d^d y/(2\pi\hbar)^d\,\,\,, \label{xp} %
\end{eqnarray}
where\, $d$ is number of pairs $\{q,p\}$, i.e. degrees of freedom %
of the dynamical subsystem (DS). %
The function-operator $\rho(t,x,p)$ is jointly the Wigner's %
probability distribution  function of DS's variables, - %
with $x$ representing the coordinates, - %
and density matrix of the thermostat. %

From viewpoint of this distribution, evidently,
\begin{eqnarray}
\rho(t,x,y)\,=\, \int \exp{(ipy/\hbar)}\, %
\rho(t,x,p)\, d^d p\,\,  \label{yp}
\end{eqnarray}
is nothing but characteristic function of the momenta $p$, %
in the same sense of the words ``characteristic function'' (CF) %
as in the probability theory. %
Or, to be more precise, $\rho(t,x,y)$ is a hybrid %
of  CF, - in respect to DS's momenta, - and probability %
distribution (PD), or density matrix, in respect to %
DS's coordinates and all thermostat's variables. %

{\bf 2}.\, %
Next, let us make transform  %
\begin{eqnarray}
\rho(t,x,y)\,=\, \exp{(-ix\Pi/\hbar)}\, R(t,x,y)\, %
\exp{(ix\Pi/\hbar)}\,\,\,, \label{r}\\ %
\rho(t,x,p)\,=\, \exp{(-ix\Pi/\hbar)}\, R(t,x,p)\, %
\exp{(ix\Pi/\hbar)}\,\,\, \nonumber %
\end{eqnarray}
After it, in terms of new (distribution) function-operator %
(density matrix) $R(t,x,y)$, Eq.\ref{e} takes form
\begin{eqnarray}
\dot{R}\, =\, %
\frac {i\hbar}m \, \frac {\pa ^2 R}{\pa x \,\pa y}\,+ %
 \,\mathcal{L}_f\,R\, =\,  %
\mathcal{L}\,R\,\,\,,\, \label{er}\\
\mathcal{L}_f\,R\, \equiv\, %
\frac i\hbar\, f\cdot y\,R\, + %
\frac 1m\, \frac {\pa }{\pa y}\, [\Pi,R]\, + %
\frac 1{i\hbar}\, %
\{\,H_b(y/2)\,R\,-\,R\,H_b(-y/2)\,\}\, %
\,\,\, \,\,\, \label{lf} %
\end{eqnarray}
Clearly, one can say that $R$ describes thermostat %
in movable frame connected to DS coordinates $q$ %
(via relative coordinates, if say in classical language). %

One more Fourier transform, in addition to (\ref{yp}), %
in respect to $x$, produces function-operator %
\begin{eqnarray}
R(t,k,y)\,=\, \int \exp{(ikx)}\, %
R(t,x,y)\, d^dx \,\,\,, \label{ky} %
\end{eqnarray}
which is joint semiclassical CF of all DS's %
variables ``hybridized'' with thermostat's density matrix.
The corresponding evolution equation directly %
follows from Eq.\ref{er}\,:
\begin{eqnarray}
\dot{R}\, =\, %
\frac {\hbar\,k}m \, \frac {\pa R}{\pa y}\,+ %
\mathcal{L}_f\,R\, %
\equiv\, \mathcal{L}\,R\,\,\, \label{eky} %
\end{eqnarray}

{\bf 3}.\, %
For last step, we assume that, - %
as usually in many-particle physics (see e.g. \cite{mah}), - %
all possible states of the thermostat can be described %
as results of actions of definite particles %
(or quasi-particles) and/or quanta creation %
and annihilation operators, $c^\dagger_s $ and $c_s$, %
onto  ``vacuum state''. Index\, ``s''\, here enumerates %
various sorts and modes of particles and quanta. %
Then all thermostat's observables, including %
$H_b(q)\,$ and $\,\Pi\,$, can be treated as functions %
of $c^\dagger_s $ and $c_s$. Correspondingly, %
thermostat's component of full  density matrix %
of our system can be completely described in terms %
of quantum CF defined e.g. like %
$\, \Tr_B\, \exp{(z^*_s\,c_s)}\, \rho\, %
\exp{(z_s\,c_s^\dagger)}\,$\,, %
where $\Tr_B$ denotes trace over thermostat space,\, %
the repeated indices ``s'' mean summation %
(or/and integration),\, and $z_s$ and $z_s^*$ %
are test, or probe, parameters, either complex %
$c$-numbers or Grassmann numbers, %
depending on whether $c^\dagger_s $, $c_s$  %
are Bose or Fermi operators %
(at that, generally, $z_s$ and $z_s^*$ are thought %
as independent, not mutually conjugated, variables). %

Then, analogously, let us introduce full CF of our system by
\begin{eqnarray}
\mathcal{F}\{t,k,y,\,z,z^*\}\,=\, %
\Tr_B\, \exp{(z^*_s\,c_s)}\, R(t,x,y)\, %
\exp{(z_s\,c_s^\dagger)}\,\,\, \label{qcf} %
\end{eqnarray}

Importantly, under the coherent-state representation of $R$ %
such defined CF coincides %
with mere CF of corresponding ``quasi-probability density''. %
Therefore $\mathcal{F}\{t,k,y,\,z,z^*\}$ is fully %
semi-classical object, and we can speak about it %
applying usual ``probability-theoretical'' terminology.

In order to write out an evolution equation for it, %
first for any operator $A$, - composed of %
various $c^\dagger_s $ and $c_s$, - let us introduce two %
operators $\mathcal{A}^+$ and $\mathcal{A}^-$ %
acting in space of functions of the probe parameters %
and defined as follow:
\begin{eqnarray}
\exp{(z_s\,c_s^\dagger)}\, %
\exp{(z^*_s\,c_s)}\,A\, =\, \nonumber\\ =\,
\mathcal{A}^+\left(z,z^*, \frac {\pa}{\pa z}, %
\frac {\pa}{\pa z^*}\right)\, %
\exp{(z_s\,c_s^\dagger)}\, %
\exp{(z^*_s\,c_s)} \,\,\,, \,\,\,\,\,\,\label{ap}\\ %
A\, \exp{(z_s\,c_s^\dagger)}\, %
\exp{(z^*_s\,c_s)}\, =\, \nonumber\\ =\,
\mathcal{A}^-\left(z,z^*, \frac {\pa}{\pa z}, %
\frac {\pa}{\pa z^*}\right)\, %
\exp{(z_s\,c_s^\dagger)}\, %
\exp{(z^*_s\,c_s)} \,\,\,  \,\,\,\,\,\,\label{am} %
\end{eqnarray}
Thus $\mathcal{A}^+$ and $\mathcal{A}^-$ %
are composed of multiplications and differentiations  %
in respect to probe parameters. %
Obviously, structure of $\mathcal{A}^+$ and $\mathcal{A}^-$ %
is unambiguously determined by that of $A$ and %
commutation rules of $c^\dagger $\,'s and $c$\,'s. %
At that, anyway
\begin{eqnarray}
\left[ \mathcal{A}^+\left(z,z^*, \frac {\pa}{\pa z}, %
\frac {\pa}{\pa z^*}\right)\,-\, %
\mathcal{A}^-\left(z,z^*, \frac {\pa}{\pa z}, %
\frac {\pa}{\pa z^*}\right)\, %
\right]_{\,z=z^*=0}\,=\,0\,\,\, \label{id} %
\end{eqnarray}

Exploiting these definitions, one easy transforms Eq.\ref{eky} %
into equation for the full CF $\mathcal{F}\{t,k,y,\,z,z^*\}$\,: %
\begin{eqnarray}
\dot{\mathcal{F}}\, =\, %
\left[\, \frac {\hbar\,k}m \, %
\frac {\pa }{\pa y}\,+ \, %
\mathcal{L}_f\,\right]\, \mathcal{F}\, %
=\, \mathcal{L}\,\mathcal{F}\,\,\,, \label{ecf}\\
\mathcal{L}_f\, =\, %
\frac {i\,f\cdot y}{\hbar}\, +\, %
\frac 1m\, \frac {\pa }{\pa y}\, %
[\,\mathcal{P}^+\,-\,\mathcal{P}^- \,]\, +\, %
\frac 1{i\hbar}\, %
[\,\mathcal{H}_b^+(y/2)\,-\, %
\mathcal{H}_b^-(-y/2)\,]\,\,\, \,\,\,\, \label{l0} %
\end{eqnarray}
Here\ $\mathcal{H}_b^\pm(q)$\, and %
$\mathcal{P}^\pm$\, %
are the operators produced, - according to %
Eqs.\ref{ap}-\ref{am}\,, - from %
$H_b(q)$ and $\Pi$, respectively, and their %
arguments are omitted for brevity. %

Further, let us discuss construction and properties %
of solutions of Eq.\ref{ecf} and equivalent Eqs.\ref{er} and \ref{eky}.

\section{Equilibrium and  steady non-equilibrium states}

First, consider stationary solutions of Eq.\ref{ecf}, %
\ref{eky}, \ref{er} and \ref{e}. %
Because of the thermostat's %
property (\ref{p}) we can expect that stationary solutions %
of Eq.\ref{er} are independent on $x$\,, %
i.e. ``spatially uniform'', if one interprets %
the DS as ``Brownian particle'' (BP), while the thermostat %
as a medium where BP moves %
(then property (\ref{p}) means thermostat's spatial infiniteness %
and uniformity). %
Correspondingly,  %
Eqs.\ref{eky} and \ref{ecf} have (non-trivial) %
stationary solutions only when the %
parameter $k$ (``wave vector'') equals to zero, $k=0$. %
Denoting them by %
$R_{st}(y;f)$\, and\ %
$\mathcal{F}_{st} =\mathcal{F}_{st}\{y,z,z^*; f\}$, %
we can write\,
\begin{eqnarray}
\mathcal{L}_f\, R_{st}\,=\,0\, \,\,, %
\,\,\,\,\, 
\mathcal{L}_f\, \mathcal{F}_{st}\, %
=\,0\,\,\,, \label{st}
\end{eqnarray}
with\, $\mathcal{L}_f$\, defined by (\ref{lf}) %
and (\ref{l0}), respectively.

Among all possible solutions of this equation, %
most physically interesting are that characterized %
by definite thermostat's  temperature. %
In absence of the ``external driving force'', %
i.e. at $f=0$\,, such solution is thermodynamically %
equilibrium one determined by  %
\begin{eqnarray}
R_{eq}(y)=\rho_{eq}(y)\,\propto\, %
\langle y/2|\, \exp{\{-[p^2/2m +H_b(q)]/T\}}\, %
|-y/2\rangle \,\,\,, \label{ye}\\
\mathcal{F}_{eq}\{y,\,z,z^*\}\,=\, %
\Tr_B\, \exp{(z^*_s\,c_s)}\, R_{eq}(y)\, %
\exp{(z_s\,c_s^\dagger)}\,\,\,, \label{cfe} %
\end{eqnarray}
along with obvious normalization condition\, %
$ \Tr_B\, R_{eq}(0)\,=1$\,, %
$\mathcal{F}_{eq}\{0,0,0\}=1$\,. %

Further, let us take these equilibrium expressions to be %
initial conditions to Eqs.\ref{eky} and \ref{ecf} %
with\, $f=\,$const\,$\neq 0$\, and\, $k=0$\,. %
Then stationary asymptotic of their solutions at $t\rightarrow\infty$ %
will give us such non-equilibrium %
solution of Eq.\ref{st} which can be treated as %
perturbation of $R_{eq}$ and $\mathcal{F}_{eq}$ and therefore  also %
is characterized by definite thermostat temperature. %

\section{Characteristic functional, statistical correlations, %
cumulants, and statistics of the Brownian motion}

Of course, we just have made second principal assumption, %
namely, that construction of the Hamiltonian\, $H_b(q)$\, ensures %
existence of the mentioned stationary asymptotic, %
at least at sufficiently small finite $\,|f|$\,. %

If it is so, then we can investigate stationary random walk %
of the BP, - i.e. random changes of the DS variable\, $q(t)$\,, - %
by considering non-stationary solutions of %
Eqs.\ref{er} or \ref{eky} or \ref{ecf}  %
at $t>0$\, with initial conditions
\begin{eqnarray}
R(t=0,x,y)\,=\,\delta(x)\,R_{st}(y;f)\,\,,\,\,\,\, %
R(t=0,k,y)\,=\,R_{st}(y;f)\,\,,\,\,\, \nonumber\\ %
\mathcal{F}\{t=0,k,y,z,z^*\}\,=\, %
\mathcal{F}_{st}\{y,z,z^*;f\}\,\,, \label{ic}
\end{eqnarray}
respectively. %
Evidently, these initial conditions are quasi-classical %
representations of such  state of DS (BP) which is balanced %
in respect to its interaction with thermostat but, at the same time, %
localized in the\, $q\,$-space at $q(t=0)=0$. %
Therefore solution of Eq.\ref{er} for $t>0$\, determines %
probability density distribution,\, $W(t,x)$\,, %
of\, $x(t)=q(t)-q(0)$\, (BP's path, or displacement, %
during time $t$), while Eqs.\ref{eky} and \ref{ecf}  %
characteristic function of this distribution: %
\begin{eqnarray}
W(t,x)= \Tr_B\,R(t,x,y=0)\,=\, %
\langle \delta(x(t)-x)\rangle\,\,,\,\, \label{wx}\\  %
W(t,k)=\int e^{ikx}\,W(t,x)\,d^dx\,=\, %
\nonumber\\ =\,
\Tr_B\,R(t,k,y=0)\,=\, %
\mathcal{F}\{t,k,y=0,z=0,z^*=0\}\,=\, %
\langle \,e^{ikx(t)}\,\rangle\,\, \label{wk}
\end{eqnarray}

We introduced the angle brackets as standard comfortable %
designation of statistical ensemble averaging. With this %
designation, in the quasi-classical language, we can write
\begin{eqnarray}
\mathcal{F}\{t,k,y,z,z^*\}\,=\, %
\,\left\langle\, \exp{[\,ikx(t)+ %
i\xi v(t)\, + z_s c^*_s(t) + %
z_s^* c_s(t)\, ]} %
\,\right\rangle\,\,, \label{av}
\end{eqnarray}
where\, $v(t)=p(t)/m =dx(t)/dt\,$\, represents velocity of %
$q(t)\,$'s changes, so that
\begin{eqnarray}
x(t)\,= \int_0^t v(\tau)\,d\tau\,\,\,, \label{int}
\end{eqnarray}
and\,\, $\xi\,\equiv\, my/\hbar\,$. %

Expression (\ref{av}) visually shows %
that\, $\,F\{t,k,y,z,z^*\}\,$\, %
is full characteristic function of all variables of the whole %
system ``DS plus thermostat'' (or, to be be precise, %
characteristic functional, since for infinitely large thermostat %
$\,c_s$\, and \, $c_s^*$\, form continuum set of variables). %
Taking in mind general properties of characteristic functions %
in the probability theory \cite{fel}, %
we can write also
\begin{eqnarray}
F\{t,k,y,z,z^*\}\,=\,W(t,k)\,\, %
\Theta\{t,k,y,z,z^*\}\,\, %
\mathcal{F}_{st}\{y,z,z^*;f\}\,\,\,, \label{ex}
\end{eqnarray}
where the middle multiplier on the right, %
$\,\Theta\,$, contains all cross-correlations %
between the path $\,x(t)\,$, on one hand, and  %
$\,v(t)\,$, $\,c_s(t)$\, and \, $c_s^*(t)$\,, %
on the other hand. This means that %
\begin{eqnarray}
\Theta\{t,k=0,y,z,z^*\}\,=\, %
\Theta\{t,k,y=0,z=0,z^*=0\}\,=\,1\,\,\, \label{th}
\end{eqnarray}
In terms of cumulants, i.e. irreducible correlations, - %
to be designated by double angle brackets, - %
\begin{eqnarray}
\ln\, \Theta\{t,k,y,z,z^*\}\,=\, %
\sum_{n,l\,=\,1}^\infty \, \frac {(ik)^n} %
{n!\,l!} \,  %
\left\langle\left\langle\,x^n(t)\,\eta^l(t)\, %
\right\rangle\right\rangle\, =\, %
\, \nonumber\\ \,=\, %
\sum_{n,l\,=\,1}^\infty \, \frac {(ik)^n} %
{n!\,l!} \, \int_0^t \dots \int_0^t  %
\left\langle\left\langle\,v(\tau_1)\dots %
v(\tau_n)\,\eta^l(t)\, %
\right\rangle\right\rangle\, %
d\tau_1\dots d\tau_n \,\,\,, \,\,\, \label{thc}\\
\ln\, W(t,k)\,=\,%
\sum_{n\,=\,1}^\infty \, \frac {(ik)^n} %
{n!} \,  %
\left\langle\left\langle\,x^n(t)\, %
\right\rangle\right\rangle\,\,\,, \,\,\, \label{wkc}
\end{eqnarray}
where\, $\eta(t)= i\xi v(t) + z_s c^*_s(t) + %
z_s^* c_s(t)\,$\,. %

Before discussion of possible time behavior of %
the\, $x(t)\,$'s cumulants in (\ref{wkc}) and thus %
statistics of the  Brownian motion, %
we have to realize most principal property of the evolution  %
operator $\mathcal{L}$ in Eq.\ref{ecf} determining %
the\, $x(t)\,$'s cumulants. Namely, to take into account that

\section{Spectrum of the Liouville super-operator %
is purely imaginary}

Let\, $\Psi_\alpha\,$ is complete set of mutually orthogonal %
eigen-vectors of the full system's Hamiltonian (\ref{h}): %
$\,H|\Psi_\alpha\rangle = E_\alpha |\Psi_\alpha\rangle \,$. %
Then\,
$\,\rho_{\alpha\beta}\,=\, %
|\Psi_\alpha\rangle\langle \Psi_\beta |\,$ %
is complete set of orthogonal %
eigen-vectors of the Liouville, or von Neumann, %
super-operator\, $\mathcal{L}$\, in (\ref{e}): %
$\, \mathcal{L}\rho_{\alpha\beta} = %
i((E_\beta -E_\alpha )/\hbar )\rho_{\alpha\beta} \,$. %
The orthogonality is understood in the sense of %
the natural ``scalar product'' %
$\,\,(A,B)\,=\,\Tr\, A^\dagger B\, %
=\, \Tr_D\,\Tr_B\,A^\dagger B\, $\,. %

Hence, $\,\mathcal{L}\,$ has purely imaginary spectrum. %
This statement equally comprises, of course, %
the evolution operators in Eqs.\ref{er}, \ref{eky} and %
\ref{ecf} which are equivalent representations of %
original Liouville operator. %
The same can be said about the operators $\,\mathcal{L}_f\,$. %

In other words, these operators can not have real eigen-values %
or complex ones with nonzero real parts. %
This trivial truth of statistical mechanics %
is sufficient ground for fundamental theorem to be %
claimed in next section.

Before it, notice, first, that in the representation defined by %
Eqs.\ref{r} and \ref{er} the mentioned scalar product reads %
\begin{eqnarray}
(A,B)\,=\, \int\int \Tr_B\, %
A^\dagger(x,-y)\,B(x,y)\, d^dy\,d^dx\,\,\, \label{spr}
\end{eqnarray}
In the representations producing Eqs.\ref{eky} %
and \ref{ecf}, evidently, the variable (``wave vector'') %
$\,k\,$ in fact plays role of passive parameter %
of the Liouville super-operator $\,\mathcal{L}\,$. %
Therefore   its eigen-vectors naturally %
divide into layers (subsets) which correspond %
to different $\,k\,$'s and can be marked by $\,k\,$  %
(such treatment of $\,\mathcal{L}\,$'s eigen-values %
will appear below). %
At that, the scalar product (\ref{spr}) reduces, - %
in case of Eq.\ref{eky}, - to  %
\begin{eqnarray}
(A,B)\,=\, \int \Tr_B\, %
A^\dagger(k,-y)\,B(k,y)\, d^dy\,\,\, \label{spky}
\end{eqnarray}
in each separate layer (an equivalent formula %
for the case of Eq.\ref{ecf} looks rather cumbersome %
and will be written out elsewhere).

\section{Theorem about uncertainty of relaxation, diffusion and %
dissipation rates}

The general hope of conventional ``kinetic theory'', or %
``physical kinetics'', etc., %
is that a proper interaction of DS (a part of much greater system) %
with thermostat (the rest of the system) can impel %
the first of them to irreversible and stochastic behavior %
characterized by well definite ``kinetic coefficients'', %
relaxation and dissipation rates, etc., %
at least under the ``thermodynamic limit'' (for infinitely %
large thermostat). %
In respect to presently considered class of systems, %
this assumption means that all the cumulant functions %
in (\ref{thc}) and (\ref{wkc}) are fast enough decaying %
(integrable) functions of time %
differences $\,t-\tau_j\,$  %
and $\,\tau_j-\tau_k\,$, respectively. %
That is all terms %
of series (\ref{thc}) tend with time to finite limits, so that %
\begin{eqnarray}
\Theta\{t,k,y,z,z^*\}\,\rightarrow\, %
\Theta\{\infty,k,y,z,z^*\}\, %
\neq\,\,\infty\,,\,0\,\,\,,\,\,  \label{thl}
\end{eqnarray}
while all terms of series (\ref{wkc}) are asymptotically linear  %
time functions:
\begin{eqnarray}
\ln\, W(t,k)\,\rightarrow\, %
\lambda(ik)\,t\,+\,\texttt{const}\,\,\,,\,\, \label{as}\\
\lambda(ik)\,=\, \sum_{n=1}^\infty (ik)^n %
\lambda_n(f)/n!\,=\, ik\,\langle v \rangle\, +\, %
(ik)^2\,D\,+\,\dots\,\,\,, \,\,\, \label{l}\\
\lambda_n(f)\,=\,\lim \,\, \langle\langle \,x^n(t)\, %
\rangle \rangle/t\,\,\,, \,\,\label{ll}
\end{eqnarray}
where\, $\langle v \rangle = \langle dx(t)/dt \rangle %
= \langle x(t)/t \rangle\,$ is mean velocity vector %
and\, $D\,$ is diffusivity tensor. For ``good'' enough %
thermostat and small force $\,f\,$ one expects also %
that $\,\langle v \rangle =\mu f\,$ with %
$\,\mu $ being mobility tensor. %

In terms of the probability theory \cite{fel}, %
Eq.\ref{as} says that the Brownian path $\,x(t)\,$ %
behaves as a random process with %
independent increments and infinitely divisible probability %
distribution, and Eq.\ref{l} specifies that this is diffusive %
process with asymptotically Gaussian probability %
distribution. %

But, fortunately, these conventional assumptions %
can not be true. Indeed, if they were true then expressions %
(\ref{thl})-(\ref{as}) as combined with Eqs.\ref{ecf} and \ref{ex} %
would imply that asymptotically
\begin{eqnarray}
\lambda(ik)\,\, \Theta\{\infty,k,y,z,z^*\}\, %
\mathcal{F}_{st}\{y,z,z^*;f\}\, =\, %
\nonumber\\ \,=\, %
\mathcal{L}\,\, \Theta\{\infty,k,y,z,z^*\}\, %
\mathcal{F}_{st}\{y,z,z^*;f\}\,\,\,, \, \label{asl}
\end{eqnarray}
which would means that\, $\lambda(ik)\,$  is eigen-value of the %
Liouville super-operator. But it is certainly impossible %
since\, $\lambda(ik)\,$ by its definition %
inevitably has non-zero (negative) real %
component,\, $\Re\,\lambda(ik)< 0\,$ at $\,k\neq 0\,$\, %
(while, to be recalled, $\,\mathcal{L}\,$'s spectrum is purely imaginary!). %

Consequently, the hypothetical linear asymptotic (\ref{as})-(\ref{ll}) %
is wrong, and in fact
\begin{eqnarray}
\frac {\langle\langle \,x^n(t)\, \rangle \rangle}t\, %
\rightarrow\,\infty \,\, \,\, \label{sl}
\end{eqnarray}
for some $\,n>1\,$ (or $\,n>3\,$ at $\,f=0\,$), %
that is all particular increments of the path $\,x(t)\,$ %
are essentially statistically dependent one on another %
(correlated with each other) %
regardless of time distance between them. %

At that, values $\,n=1\,$, or $\,n<3\,$ at $\,f=0\,$, %
are excluded from candidates to the super-linearity %
(as well as to sub-linearity)  by our assumptions about %
existence of steady non-equilibrium state %
(described by$\,\mathcal{F}_{st}\,$) %
with finite mean (``drift'') velocity $\,\langle v\rangle\,$, 
which means finiteness of BP's diffusivity in  equilibrium state %
(described by $\,\mathcal{F}_{eq}\,$) at $\,f=0\,$ %
(this follows from the Einstein relation $\,D=T\mu\,$ %
easy provable for our systems too).

Therefore, the crash of the ``independence of increments'' impliess %
that generally the super-linear cumulants' %
asymptotic (\ref{sl}) takes place at %
any $\,n>1\,$ (or $\,n>3\,$, if $\,f=0\,$). %
Then, it remains to realize that this statement %
is equivalent to statement %
that mobility and diffusivity possess low-frequency fluctuations %
like ``flicker noise'' or 1/f\,-noise. %

In other words, the mobility and diffusivity %
(as well as  related ``kinetic characteristics'' of $\,x(t)\,$'s %
interaction with thermostat) do not have certain values %
but change from one experiment (measurement) to another, %
with a spread what anyway hugely exceeds limits suggested %
by the ``law of large numbers''. %

\section{Conclusion}

We have expounded rather general theorem %
(or, strictly speaking, a ``storage'' of future rigorous theorem) %
stating that interaction of one Hamiltonian %
(sub-) system with another, - let serving as an arbitrarily %
 large thermostat, - never can ensure well certain %
quantitative characteristics of irreversible and dissipative %
behavior of the first of them. %
Instead, these characteristics undergo significant %
1/f\,-type (``flicker'' type) low-frequency fluctuations. %
Citing once again myself from \cite{bk3}, %
``1/f-noise is a kind of tribute to be paid to the dynamics for %
dissipation and irreversibility properties %
of physical systems''. %

Particular variant of this theorem was obtained in \cite{oct}. %
For better understanding its pre-history, physical meaning %
and possible applications, see also works referred there %
and in the Introduction above. %
Detailed elaboration and concrete applications of our present result %
will be considered separately. %

\,\,\,

--------------------------------------------------

\,\,\,


\end{document}